\documentclass[submission,copyright,creativecommons]{eptcs}
\providecommand{\event}{REFINE 2018} 
\usepackage{breakurl}             
\usepackage{underscore}           

\usepackage[utf8x]{inputenc}
\usepackage{subcaption}
\usepackage{makeidx}  
\usepackage{amsmath}
\usepackage{amssymb}
\usepackage{graphicx}
\usepackage[svgnames]{xcolor}
\usepackage[version=0.96]{pgf}
\usepackage{tikz}
\usetikzlibrary{arrows,shapes,automata,backgrounds,petri,positioning}
\usepackage{bbm}
\usepackage{ucs}
\usepackage{autofe}
\usepackage{fancyvrb}
\usepackage{enumitem}
\usepackage{array,multirow}
\usepackage{multicol}
\usepackage{url}
\usepackage{xspace}

\newcolumntype{C}[1]{>{\centering\let\newline\\\arraybackslash\hspace{0pt}}m{#1}}

\DeclareUnicodeCharacter{8759}{\ensuremath{::}}
\DeclareUnicodeCharacter{8344}{\ensuremath{_m}}
\DeclareUnicodeCharacter{7523}{\ensuremath{_r}}
\DeclareUnicodeCharacter{8343}{\ensuremath{_l}}
\DeclareUnicodeCharacter{8339}{\ensuremath{_x}}
\DeclareUnicodeCharacter{8342}{\ensuremath{_k}}
\DeclareUnicodeCharacter{7522}{\ensuremath{_i}}
\DeclareUnicodeCharacter{945}{\ensuremath{\alpha}}
\DeclareUnicodeCharacter{946}{\ensuremath{\beta}}
\DeclareUnicodeCharacter{955}{\ensuremath{\lambda}}
\PrerenderUnicode{é}
\PrerenderUnicode{ê}
\PrerenderUnicode{è}
\PrerenderUnicode{ë}
\PrerenderUnicode{à}
\PrerenderUnicode{ä}
\PrerenderUnicode{â}
\PrerenderUnicode{î}
\PrerenderUnicode{ï}
\PrerenderUnicode{ö}
\PrerenderUnicode{ô}
\PrerenderUnicode{ù}
\PrerenderUnicode{ü}

\newcommand{\lustre}[0]{\textsc{Lustre}\xspace}
\newcommand{\dsml}[0]{\textsc{DSML}\xspace}
\newcommand{\moc}[0]{\textsc{MoC}\xspace}
\newcommand{\ccsl}[0]{\textsc{CCSL}\xspace}
\newcommand{\cps}[0]{\textsc{CPS}\xspace}
\newcommand{\becool}[0]{\textsc{BeCooL}\xspace}

\newcommand{\agda}[0]{Agda\xspace}
\newcommand{\coq}[0]{Coq\xspace}
\newcommand{\gemoc}[0]{\textsc{Gemoc}\xspace}

\makeatletter
\newcommand*{\storedata}[2]{%
  \count@=0 %
  \@tfor\@tmp:=#2\do{%
    \advance\count@\@ne
    \expandafter\let\csname data:\the\count@:#1\endcsname\@tmp
  }%
  \expandafter\edef\csname data:0:#1\endcsname{\the\count@}%
}
\newcommand*{\getdata}[2]{%
  \@ifundefined{data:0:#2}{%
    \@latex@error{Undefined data `#2'}\@ehc
  }{%
    \expandafter\@getdata\expandafter{%
      \the\numexpr
        \ifnum\numexpr(#1)<\z@
          \@nameuse{data:0:#2}+1+%
        \fi
        (#1)%
      \relax
    }{#2}{#1}%
  }%
}
\newcommand*{\@getdata}[3]{%
  \ifnum#1<\z@
    \@getdata@error{\the\numexpr(#3)\relax}{#2}%
  \else
    \ifnum#1>\@nameuse{data:0:#2} %
      \@getdata@error{#1}{#2}%
    \else
      \@nameuse{data:#1:#2}%
    \fi
  \fi
}
\newcommand*{\@getdata@error}[2]{%
  \@latex@error{%
    Wrong data selector #1 for `#2',\MessageBreak
    which only contains \@nameuse{data:0:#2} item(s)%
  }\@ehc
}
\makeatother

\title{Ordering Strict Partial Orders to Model\\Behavioral Refinement}

\author{Mathieu Montin
\institute{Université de Toulouse ; Toulouse INP,  \textit{IRIT} \\
2 rue Camichel, BP 7122, 31071 Toulouse Cedex 7, France}
\institute{\textit{CNRS} ; Institut de Recherche en Informatique de Toulouse (IRIT)}
\email{mathieu.montin@enseeiht.fr}
\and
Marc Pantel
\institute{Université de Toulouse ; Toulouse INP,  \textit{IRIT} \\
2 rue Camichel, BP 7122, 31071 Toulouse Cedex 7, France}
\institute{\textit{CNRS} ; Institut de Recherche en Informatique de Toulouse (IRIT)}
\email{marc.pantel@enseeiht.fr}
}
\def\titlerunning{Ordering Strict Partial Orders to Model Refinement}
\def\authorrunning{Mathieu Montin et al.}

\begin{document}
\maketitle

\begin{abstract}
Software is now ubiquitous and involved in complex interactions with the human users and the physical world in so-called cyber-physical systems (\cps) where the management of time is a major issue. Separation of concerns is a key asset in the development of these ever more complex systems. Two different kinds of separation exist: a first one corresponds to the different steps in a development leading from the abstract requirements to the system implementation and is qualified as vertical. It matches the commonly used notion of refinement. A second one corresponds to the various components in the system architecture at a given level of refinement and is called horizontal. Refinement has been studied thoroughly for the data, functional and concurrency concerns while our work focuses on the time modelling concern. This contribution aims at providing a formal construct for the verification of refinement in time models, through the definition of an order between strict partial orders used to relate the different instants in asynchronous systems. This relation allows the designer at the concrete level to distinguish events that are coincident at the abstract level  while preserving the properties assessed at the abstract level. This work has been conducted using the proof assistant \agda and is connected to a previous work on the asynchronous language \ccsl, which has also been modelled using the same tool.
\end{abstract}

\section{Introduction}

\subsection{Separation of concerns}

Nowadays, many devices require to handle complex interactions with both the human users and the physical world. These devices, like cars, aircrafts, trains, rockets, satellites, pacemakers, robots, etc are called Cyber-Physical systems (\cps). While these ones offer more and more advanced and complex services, they become increasingly dense and complex, which leads their developers to use separation of concerns throughout the different phases of their development. There exists two kinds of separations of concerns : the first one is qualified as horizontal and aims at describing complex systems through the different physical -- or logical -- parts they contain. The second one is qualified as vertical and corresponds to the commonly used notion of refinement, where the different levels of abstraction of a given systems are described separately while the properties they exhibit are preserved throughout these steps.

Horizontal separation is usually handled at design time through the expression of the various system parts in different Domain Specific Modelling Languages (\dsml). Their execution, for validation and verification purposes, may rely on different Models of Computation (\moc). A sophisticated coordination of the various events occurring in the different parts is thus needed to observe the global behavior of the system. For \cps, the modeling of time in the various \dsml and the coordination between the different time models is a major issue. This heterogenous modelling approach has been integrated in the Ptolemy toolset proposed by Lee et al.~\cite{DBLP:journals/ijcs/BuckHLM94}, the ModHel'X toolset proposed by Boulanger et al.~\cite{DBLP:conf/models/HardebolleB07} and the \gemoc studio proposed by Combemale et al.~\cite{DBLP:journals/computer/CombemaleDBFJG14}. Our work targets a proof based formal modeling and verification framework to prove properties of languages and models in such toolsets.

\indent Vertical separation usually enforces a refinement relation between the different models of the same part of the system in order to ensure the consistency of the various global executions. This approach is for example advocated by the B and Event-B methods~\cite{DBLP:books/daglib/0015096,DBLP:books/daglib/0024570,DBLP:conf/zum/AbrialCM05} in order to prove the preservation of the properties from the specification to the implementation. In the case of asynchronous systems, refinement is usually related to simulation and corresponds to replacing $\tau$ transition by effective actions. In the case of synchronous systems, refinement corresponds to decomposing an instant at a given level into several instants at the refined level. Synchronous refinement has been widely studied in the case of synchronous \moc first as oversampling for data-flow languages~\cite{MikacCaspi2005} and then as time refinement for reactive languages~\cite{DBLP:journals/ejes/GemundeBS13,DBLP:journals/scp/MandelPP15}. Polychronous time models have been used to assess the vertical refinement during system design~\cite{DBLP:journals/fuin/TalpinGSDG04}. Their relational nature is more appropriate at design time as it introduces less constraints than the common functional computation of clocks in synchronous programming languages derived from \lustre. Thus, the refinement has to be made explicit in our formal framework.

\subsection{Context}

\indent Our work focuses on the modeling of time in \gemoc that mixes both horizontal and vertical separation of concerns. Indeed, \gemoc allows to define the \dsml used to model the various parts in a \cps in each phase of their development. Thus, \dsml are combined both in an horizontal and vertical manners. \gemoc relies on the UML MARTE \ccsl (Clock Constraint Specific Language) to model both the \moc for the various \dsml~\cite{DBLP:conf/sle/CombemaleDLMBBF13,DBLP:conf/date/DeAntoniDTCC15,DBLP:conf/sle/LatombeCCDP15} and the coordination between \dsml using the Behavioral Coordination Language (\becool)~\cite{DBLP:conf/models/LarsenDCM15}. Our work in \gemoc targeted an example of horizontal separation. This contribution targets the vertical separation. More precisely, we want to assess the relations between the various time concerns in the various models of the same system part in a vertical separation of concerns. In that purpose, we provide a mechanized definition for the vertical relation and eventually apply it to \ccsl.

\indent This issue is handled by introducing an instant refinement relation inspired from time refinement in order to ultimately combine both horizontal and vertical separation of concern in the design of heterogeneous systems. In time models that depict the temporal execution of heterogeneous systems, partial orders are usually used to bind the instants together. This contribution provides a formal construct for the time refinement in these models, as an order relation between these partial orders. This relation allows the designer at the concrete level to distinguish events that are coincident at the abstract level while preserving the properties assessed at the abstract level.

\indent This relation is generic and can be applied to any system, the semantics of which relies on a set of traces. It has been mechanized with the \agda proof assistant, in order to be linked with a denotational semantics of \ccsl, that has already been mechanized using \agda. This allows assessing properties of this new relation and prove that it preserves the different \ccsl operators semantics. This contribution relies on a simple example of oversampling in a synchronous system.

\subsection{State of the art}

Refinement has been thoroughly studied \cite{DBLP:conf/icfem/ReevesS03,DBLP:journals/entcs/ReevesS08a} and implemented for many different modeling and programming concerns like data \cite{DBLP:books/cu/RoeverE1998} and algorithms (sequential \cite{DBLP:conf/rex/BackW89}, concurrent \cite{DBLP:conf/rex/Back89}, distributed, etc). Time can be represented with a single global reference clock that binds all clocks in the system together~\cite{DBLP:journals/iandc/LynchV96,DBLP:journals/tcs/Broy01}. However, since building these global clocks is usually tricky, time is more often abstracted as a partial order relation~\cite{DBLP:conf/focs/Pnueli77,DBLP:journals/iandc/LynchV95}. Refinement \cite{DBLP:journals/tcs/AbadiL91} then relies on simulation \cite{DBLP:journals/fac/He89,DBLP:journals/scp/Hesselink11} or bisimulation relations between the semantics of the more abstract and concrete system models.

Our proposal provides a mechanized refinement relation formalized in the \agda proof assistant. We target its coupling with a previous mechanization of the semantics of \ccsl in the same proof assistant. This relation can be integrated with any other concurrent languages. Formal mechanization of time models has already been done using other formal methods, for example \cite{DBLP:journals/fmsd/HaleCH93} uses Higher Order Logic in Isabelle/HOL; \cite{DBLP:conf/formats/GarnachoBF13} and \cite{DBLP:conf/tacs/Paulin-Mohring01} use the Calculus of Inductive Constructions in \coq, see \cite{DBLP:series/txtcs/BertotC04}. The use of \agda in this development is motivated by the expressiveness of the language and its underlying unification mechanism, which provides an efficient interactive proof experience that other tools might lack. More on \agda can be found in \cite{DBLP:conf/tldi/Norell09}, \cite{ar4} and \cite{DBLP:conf/lernet/BoveD08}. Although \agda differs from \coq by several aspects, both of these tools rely on the same underlying intuitionist type theory, first described in \cite{ar1} and clarified in \cite{ar2}. The paper version of \ccsl denotational semantics, which is connected to this work, can be found in \cite{deantoni:hal-01082274}. TimeSquare, the tool developed to describe \ccsl systems as well as solve constraint sets has been presented in \cite{deantoni:hal-00688590}. As for \ccsl itself, it was first presented in \cite{andre:inria-00280941}.

\subsection{Two different ways of considering refinement}

Refinement is a relation between the trace semantics of two systems, a more abstract and a more concrete one that can be assessed in two different ways. While these possibilities rely on different approaches, they are ultimately equivalent. Either the concrete system is derived from the first system in a correct by construction manner, and refinement is ensured by the a correct by construction derivation method, or both systems are provided independently and refinement is assessed relying on mapping information between both systems. 

The first approach can be considered as an accretion of events. It is advocated for example by the Event-B method. It consists in building step by step correct by construction trace generators (through an operational semantics). Each step consisting in a layer of concretization, hence a layer of refinement. At a given time in this process, only the events used in the levels already described are existing. This means that the set of events evolves throughout the development of the system. This operational view is akin, as explained, to correct-by-construction development of systems from the abstract specification to the concrete executable program. Sets of traces can then possibly be created through the different possible executions of the more concrete version of the system. The relations binding the different events in these traces are deduced from properties written on the specification and preserved through the refinement process. 

The second approach relies on building both sets of possible traces itself. Refinement can be assessed on these sets  (usually described through a partial order over the instants on which events occur) regardless of the generators of these traces. This vision require to express refinement on the traces themselves, through relations instead of functions. This approach can be applied later in the development process as it does not require the system to be built throughout a correct-by-construction methodology. The traces contain all the events of the system, regardless of the level of refinement on which they appear. This paper present a relation that has to be satisfied in order to verify refinement in such cases. This relation, rather than constraining the generator of the traces or the events themselves, constrain the partial order that bind them in each layer of refinement.

These two visions are somewhat conceptually opposed and the tools used to model and describe them differ as well. We chose to use \agda in this work. Set theory is akin to describe the second approach as it naturally embeds the operation of accretion through its axioms. In type theories, as the one on which \agda is based, subsets and union are not natural, while these tools provide the right level of expressiveness to mechanize relations between quantities. This motivates the use of \agda for this work where we remain descriptive and never actually compute the traces of events on which our relation is defined. 

\section{Time and refinement}

This section briefly introduces notions inherent to time handling in asynchronous systems, from the instants to the strict partial orders binding them in time models, then proceeds to the core of our contribution: our relation of refinement between these orders. 

	\subsection{Instants}
	
\emph{Instants} are the main concept on which concurrent languages are defined. Informally, an instant is a point in time where events can occur. It matches, to a certain extend, the common vision one has about time. However, time in asynchronous systems cannot be easily depicted as a single time-line consisting of well ordered instants. This is due to the lack of knowledge one can have regarding the execution of such systems, when it is usually impossible to know, for all events and their respective instants, whether one has happened before another. 

Another difference with our common perception of time is that several instants can be coincident, which means they "happen" simultaneously. This is the case for instance when two successive events happen so close to each other that they cannot be distinguished by a given observer. In some concurrent languages, such as \ccsl, this vision is completely embraced, since no instant can "host" more than one event. This means that two events that seem to occur simultaneously will still be carried by different instants, but these instants will be coincident. This vision is closely linked to the notion of refinement, because it assumes that there exists no ultimate level of refinement on which an observer can know everything about the behavior of a system, since two coincident instants can always be distinguished when looking close enough to the execution of the system. Our relation of refinement heavily relies on this observation. Let us name this set of instants $I$.

%
%

	\subsection{Strict partial orders}
	
As explained in the previous subsection, time cannot be seen as a single line that hosts any event occurrence. Instead, it contains a possibly infinite set of timelines that link instants that are observationally related. This means that the set of instants is not coupled with a total order but rather with a partial order that represents the knowledge the observer has of the behavior of the system. This means that each pair of instant is either:	
\begin{itemize}
\item[$\bullet$] \textit{strictly comparable}, through a precedence relation $≺$
\item[$\bullet$] \textit{equivalent}, through a coincidence relation $≈$
\item[$\bullet$] \textit{independent}, which means neither equivalent nor precedent
\end{itemize}
This is important to note that a partial order is not a single relation. It consists in two relations (the one defined above) that must fulfil certain properties, which are, as a reminder: 
\begin{itemize}
\item[$\bullet$] $≈$ is an equivalence relation
\begin{itemize}
\item $≈$ is reflexive: \hfill $∀ i ∈ I : i ≈ i$
\item $≈$ is transitive: \hfill$∀ (i,j,k) ∈ I^3 : i ≈ j \land j ≈ k \Rightarrow j ≈ k$\
\item $≈$ is symmetrical: \hfill$∀ (i,j) ∈ I^2 : i ≈ j \Rightarrow j ≈ i$
\end{itemize}
\item[$\bullet$] $≺$ is irreflexive towards $≈$: \hfill$∀ (i,j) ∈ I^2 : i ≺ j \Rightarrow ¬ (i ≈ j)$
\item[$\bullet$] $≺$ is transitive: \hfill$∀ (i,j,k) ∈ I^3 : i ≺ j \land j ≺ k \Rightarrow j ≺ k$
\item[$\bullet$] $≺$ respects $≈$: \hfill$(∀ (i,j,k) ∈ I^3 : i ≈ j \land i ≺ k \Rightarrow j ≺ k) \land (∀ (i,j,k) ∈ I^3 : i ≈ j \land k ≺ i \Rightarrow k ≺ j)$
\end{itemize}

\begin{figure}[h!]
 
\begin{subfigure}[t]{0.45\textwidth}
\centering
  \begin{tikzpicture}[axis/.style={thick, ->, >=stealth'}]
  \draw[axis] (1.5,0)  -- (7,0) node(xline)[right]{};

    \foreach \x in {2,3.5,5,6.5}
      \draw (\x cm,3pt) -- (\x cm,-3pt);

    \draw (2,0) node[above=3pt] {up};
    \draw (3.5,0) node[above=3pt] {eat};
    \draw (5,0) node[above=3pt] {sho/sin};
    \draw (6.5,0) node[above=3pt] {off};
  \end{tikzpicture}
\caption{A first possible behavior}
\label{linearmorning1}
\end{subfigure}
\begin{subfigure}[t]{0.45\textwidth}
\centering
  \begin{tikzpicture}[axis/.style={thick, ->, >=stealth'}]
  \draw[axis] (1.5,0)  -- (7,0) node(xline)[right]{};

    \foreach \x in {2,3.5,5,6.5}
      \draw (\x cm,3pt) -- (\x cm,-3pt);

    \draw (2,0) node[above=3pt] {up};
    \draw (3.5,0) node[above=3pt] {sho/sin};
    \draw (5,0) node[above=3pt] {eat};
    \draw (6.5,0) node[above=3pt] {off};
  \end{tikzpicture}
\caption{A second possible behavior}
\label{linearmorning2}
\end{subfigure}
\caption{Both possible behaviors}
\end{figure}
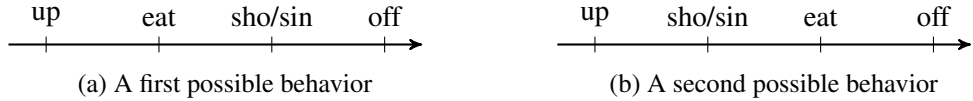

\paragraph{An example of strict partial order} Let us consider the usual morning routine of Alice. She gets up then either takes her shower first then eats or the other way around. She always sings when she showers. After that, she takes off for work. The two possible traces depicting her behavior over a single day are depicted in Figure~\ref{linearmorning1}~and~\ref{linearmorning2}. They consider the following set of possible events: getting up, showering, singing, eating and taking off, with their respecting aliases "up", "sho", "sin", "eat" and "off". 

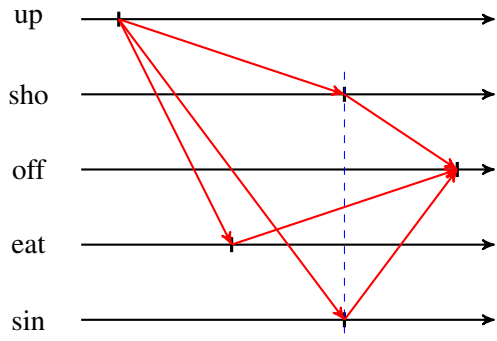
\begin{figure}[h!]
\centering
  \begin{tikzpicture}[axis/.style={thick, ->, >=stealth'}]    
    \foreach \i in {-3,-2,-1,0,1} {
      \draw[axis] (1.5,\i) -- (7,\i) node(xline)[right]{};
    }
    \draw (.8,1) node {up};
    \draw (.8,0) node {sho};
    \draw (.8,-1) node {off};
    \draw (.8,-2) node {eat};
    \draw (.8,-3) node {sin};
    
    \draw[very thick] (2 ,1.1) -- (2 ,.9);
    \draw[very thick] (5 ,0.1) -- (5 ,-.1);
    \draw[very thick] (6.5 ,-.9) -- (6.5 ,-1.1);
    \draw[very thick] (3.5 ,-1.9) -- (3.5 ,-2.1);
    \draw[very thick] (5 ,-2.9) -- (5 ,-3.1);
    
    \draw [dashed, Blue] (5 ,0.3) -- (5 ,-3.3);
    
    \draw [Red,thick,->, >=stealth'] (2,1) -- (5,0);
    \draw [Red,thick,->, >=stealth'] (5,0) -- (6.5,-1);
    \draw [Red,thick,->, >=stealth'] (2,1) -- (3.5,-2);
    \draw [Red,thick,->, >=stealth'] (3.5,-2) -- (6.5,-1);
    \draw [Red,thick,->, >=stealth'] (2,1) -- (5,-3);
    \draw [Red,thick,->, >=stealth'] (5,-3) -- (6.5,-1);
%
  \end{tikzpicture} 
  \caption{The underlying partial order}
  \label{morning}
\end{figure}

These possible behaviors are described by a time structure (derived from event structure~\cite{DBLP:journals/cacm/Lamport78,DBLP:conf/ac/Winskel86}) with an underlying partial order, that is depicted on Figure~\ref{morning}. The events "sho" and "eat" are concurrent and are not linked by any of the two relations composing the strict partial order. The blue vertical dashed line represents coincidence (when events occur simultaneously) while the red arrows represent precedence.

%
%

	\subsection{Our relation over strict partial orders}
	
These reminders about strict partial orders and instants lead to the definition of the proposed refinement relation. As our approach is part of a denotational context, we need to express a relation between certain data that are relevant to express refinement. These data cannot be the mere instants as these are not specific to a given execution, and these do not carry enough information. However, the strict partial orders binding them embed the necessary knowledge about the system behavior to be ordered in a way that respects the proposed time related instant refinement. Thus, we propose to instantiate these so-called data with the orders binding the instants together at a given level of observation. This binding of orders between instants and not instants themselves is the core contribution of this paper. The following relation takes two strict partial orders and states what it means for them to be in a relation of refinement.

\vspace{.2cm}

\hspace{-.5cm}
\fbox{\parbox{.95\columnwidth}{
Let $\Omega$ be the set of all sets : $\forall I\in\Omega, \forall (<_c, <_a, \approx_c , \approx_a)\in(I \times I)^4:\\ $
$(<_c,\approx_c) <_r (<_a,\approx_a) \overset{d}{\Longleftrightarrow} \forall (i_1,i_2) \in I:$
\vspace{-.3cm}
\[
\begin{array}{lll}
       & i_1 <_c i_2 \Rightarrow i_1 <_a i_2 \vee i_1 \approx_a i_2 & (1) \\
\wedge & i_1 <_a i_2 \Rightarrow i_1 <_c i_2 & (2) \\
\wedge & i_1 \approx_c i_2 \Rightarrow i_1 \approx_a i_2 & (3) \\
\wedge & i_1 \approx_a i_2 \Rightarrow i_1 \approx_c i_2 \vee i_1 <_c i_2 \vee i_2 <_c i_1 & (4)
\end{array}
\]
\vspace{-.3cm}
}}

\vspace{.2cm}

In this definition, the level annotated by the index $c$ is the lower (the more concrete) level of observation and $a$ is the higher (the more abstract). We state what it means for a pair of relations to refine another pair of relations. We can only compare pairs of relations that are bounded to the same underlying set. This relation is composed of four predicates, each of which indicates how one of the four relations is translated into the other level of observation. 
\begin{itemize}
\item[$\bullet$] \textit{Precedence abstraction:} If $a$ strictly precedes $b$ in the lower level, then it can either be equivalent to it in the higher level or still precede it. This means that a distinction which is visible at a lower level can either disappear at a higher level or remain visible, depending on the behavior of the refinement around these instants. 
\item[$\bullet$] \textit{Precedence embodiment:} If $a$ strictly precedes $b$ in the higher level, then it can only still precede it in the lower level. This means that the distinction between these instants was already existing in the higher level, and cannot be lost when refining. Looking closer to a system preserves precedence between instants. 
\item[$\bullet$] \textit{Coincidence abstraction:} If $a$ is equivalent to $b$ in the lower level, it can only stay equivalent in the higher level. This means that looking at the system from a higher point of view cannot reveal temporal distinction between events. 
\item[$\bullet$] \textit{Coincidence embodiment:} If $a$ is equivalent to $b$ in the higher level then the only thing we ensure is that these two instants are still related in the lower level. This means that both instants will still be related -- they cannot become independent -- but there is no guarantee on the nature of this relation.
\end{itemize}

This definition is coherent with \ccsl point of view where instants can only hold one event. Two instants appearing coincident in a given level of refinement can potentially always be refined up to a point where a distinction appears, which justifies the fact that they should not be attached to the same physical instant. 

%

\section{A refinement example}

\begin{figure}[h]
  \centering
\begin{tikzpicture}[->,>=stealth',shorten >=1pt,auto,node distance=2.8cm,semithick]
  \node[initial,state] (A)              {Off};
  \node[state]         (B) [right of=A] {On};

  \path (A) edge [bend left] node {Switch on}  (B)
        (B) edge [bend left] node {Switch off} (A)
        (B) edge [loop right] node {Execute} (B);
\end{tikzpicture}
\caption{A simple system}
\label{automaton1}
\end{figure}
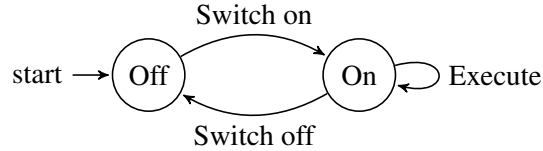

\begin{figure}[h!]
\centering
  \begin{tikzpicture}[axis/.style={thick, ->, >=stealth'}]
  \draw[axis] (1.5,0)  -- (9.7,0) node(xline)[right]{};

    \foreach \x in {2,2.9,3.8,4.7,5.6,6.5,7.4,8.3,9.2}
      \draw (\x cm,3pt) -- (\x cm,-3pt);

    \draw (2,0) node[above=3pt] {$t_\text{on}$};
    \draw (2.9,0) node[above=3pt] {$t_\text{off}$};
    \draw (3.8,0) node[above=3pt] {$t_\text{on}$};
    \draw (4.7,0) node[above=3pt] {$t_\text{ex}$};
    \draw (5.6,0) node[above=3pt] {$t_\text{ex}$};
    \draw (6.5,0) node[above=3pt] {$t_\text{off}$};
    \draw (7.4,0) node[above=3pt] {$t_\text{on}$};
    \draw (8.3,0) node[above=3pt] {$t_\text{ex}$};
    \draw (9.2,0) node[above=3pt] {$t_\text{off}$};
  \end{tikzpicture}
\caption{A trace on a single timeline}
\label{timeline1}
\end{figure}
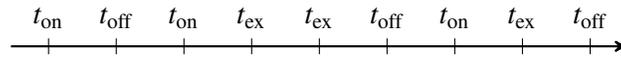

The mathematical relation defined above aims at providing a formal construct to verify refinement between traces of execution. To illustrate its relevance, we propose to apply it to a simple example chosen for its simplicity and accuracy with respect to the idea of refinement. This is a simple system whose behavior is represented as a transition system depicted on Figure~\ref{automaton1}. This system can be switched on and off. While it is on, an action can be executed any number of times. A possible trace -- amongst an infinite number of them -- of this system is depicted in Figure~\ref{timeline1}. $t_{\text{on}}$, $t_{\text{off}}$ and $t_{\text{ex}}$ respectively represent the occurrence of the "switch on", "switch off" and "execute" transitions.

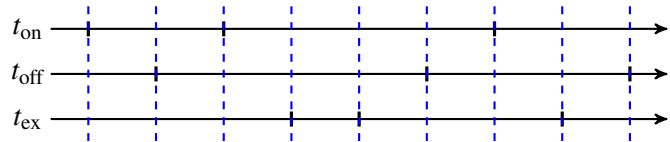
\begin{figure}[h!]
\centering
  \begin{tikzpicture}[axis/.style={thick, ->, >=stealth'}]    
    \foreach \i in {1,...,3} {
      \draw[axis] (1.5,\getdata{\i}{y}) -- (9.7,\getdata{\i}{y}) node(xline)[right]{};
    }
    \foreach \i in {1,...,3} {
      \draw[very thick] (\getdata{\i}{xOne} ,\getdata{1}{y}-0.1) -- (\getdata{\i}{xOne} ,\getdata{1}{y}+0.1);
    }
    \foreach \i in {1,...,3} {
      \draw[very thick] (\getdata{\i}{xTwo} ,\getdata{2}{y}-0.1) -- (\getdata{\i}{xTwo} ,\getdata{2}{y}+0.1);
    }
    \foreach \i in {1,...,3} {
      \draw[very thick] (\getdata{\i}{xThree} ,\getdata{3}{y}-0.1) -- (\getdata{\i}{xThree} ,\getdata{3}{y}+0.1);
    }
    \draw (1.2,\getdata{1}{y}) node {$t_{\text{on}}$};
    \draw (1.2,\getdata{2}{y}) node {$t_{\text{off}}$};
    \draw (1.2,\getdata{3}{y}) node {$t_{\text{ex}}$};

    \foreach \i in {1,...,9} {
      \draw [thick, dashed, Blue] (\getdata{\i}{dots},\getdata{1}{y}+0.3) -- (\getdata{\i}{dots},\getdata{3}{y}-0.3);
    }
  \end{tikzpicture} 
  \caption{One timeline per event}
  \label{timeline2}
  
\end{figure}

This trace starts with the birth of the system and possibly goes on indefinitely, which makes this representation partial. In addition, this design places each event on the same timeline, thus ignoring horizontal separation. In order to make it visible, we will represent, from now on, every different event on a specific timeline, such as on Figure \ref{timeline2}. This approach is used in \ccsl, where each timeline is represented by a clock which tracks the occurrences of a specific event. The instants on each timeline are totally ordered and those in the same vertical dashed blue lines are coincident.

\storedata{y}{{1.2}{.6}{0}{-.6}{-1.2}}
\storedata{xOne}{{2}{3.8}{7.4}}
\storedata{xTwo}{{2.9}{6.5}{9.2}}
\storedata{xThree}{{4.7}{5.6}{8.3}}
\storedata{xFour}{{2}{2.9}{3.8}{5.6}{6.5}{7.4}{9.2}}
\storedata{xFive}{{4.7}{8.3}}
\storedata{dots}{ {2}{2.9}{3.8}{4.7}{5.6}{6.5}{7.4}{8.3}{9.2}}

\begin{figure}[h]
  \centering
\begin{tikzpicture}[->,>=stealth',shorten >=1pt,auto,node distance=2.8cm,
                    semithick]
  \node[initial,state] (A)              {Off};
  \node[state]         (B) [right of=A] {On};

  \path (A) edge [bend left] node {$t_\text{on}$ $\{ x \leftarrow 0 \}$}  (B)
        (B) edge [bend left] node {$t_\text{off}$ $\{ x \leftarrow 0 \}$ } (A)
        (B) edge [loop right] node {$t_\text{ex}$ $\{ x \leftarrow 1 - x \}$} (B);
\end{tikzpicture}
\caption{The system pilots a light}
\label{automaton2}
\end{figure}
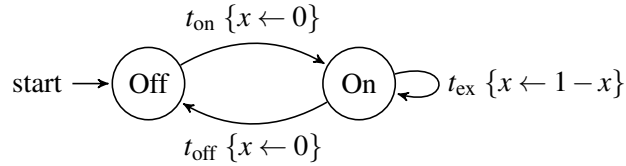

The action executed by the system while running can be specified in various ways. We imagine here that our system is connected to a light through the use of a memory containing a variable $x$. This variable is assigned by our system to the values $1$ or $0$, and the light is turned on and off accordingly. When the system is switched on, the light remains down until a button is pressed which turns it on. Pressing the same button will alternatively turn it off and on. Shutting down the system turns it off. This behavior is depicted on Figure \ref{automaton2}. 

\begin{figure}[h!]
\centering
  \begin{tikzpicture}[axis/.style={thick, ->, >=stealth'}]
    \foreach \i in {1,...,5} {
      \draw[axis] (1.5,\getdata{\i}{y}) -- (9.7,\getdata{\i}{y}) node(xline)[right]{};
    }
    \foreach \i in {1,...,3} {
      \draw[very thick] (\getdata{\i}{xOne} ,\getdata{1}{y}-0.1) -- (\getdata{\i}{xOne} ,\getdata{1}{y}+0.1);
    }
    \foreach \i in {1,...,3} {
      \draw[very thick] (\getdata{\i}{xTwo} ,\getdata{2}{y}-0.1) -- (\getdata{\i}{xTwo} ,\getdata{2}{y}+0.1);
    }
    \foreach \i in {1,...,3} {
      \draw[very thick] (\getdata{\i}{xThree} ,\getdata{3}{y}-0.1) -- (\getdata{\i}{xThree} ,\getdata{3}{y}+0.1);
    }
    \foreach \i in {1,...,7} {
      \draw[very thick] (\getdata{\i}{xFour}, \getdata{4}{y}-0.1) -- (\getdata{\i}{xFour}, \getdata{4}{y}+0.1);
    }
    \foreach \i in {1,2} {
      \draw[very thick] (\getdata{\i}{xFive}, \getdata{5}{y}-0.1) -- (\getdata{\i}{xFive}, \getdata{5}{y}+0.1);
    }
    
    \draw (1.2,\getdata{1}{y}) node {$t_{\text{on}}$};
    \draw (1.2,\getdata{2}{y}) node {$t_{\text{off}}$};
    \draw (1.2,\getdata{3}{y}) node {$t_{\text{ex}}$};
    \draw (1.2,\getdata{4}{y}) node {$t_{x_0}$};
    \draw (1.2,\getdata{5}{y}) node {$t_{x_1}$};

    \foreach \i in {1,...,9} {
      \draw [thick, dashed, Blue] (\getdata{\i}{dots},\getdata{1}{y}+0.3) -- (\getdata{\i}{dots},\getdata{5}{y}-0.3);
    }
  \end{tikzpicture} 
   \caption{The trace of the system with the addition of the variable $x$}
   \label{timelinewithx}
\end{figure}
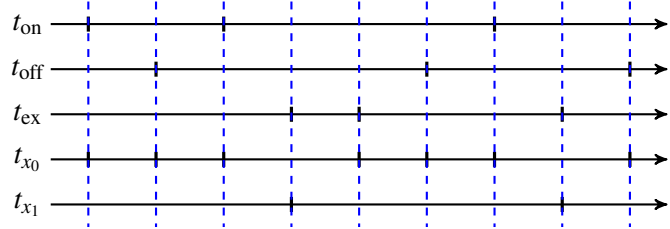

By specifying our system behavior, we defined events that can be added to its traces. $t_{x_0}$ and $t_{x_1}$ respectively correspond to the variable $x$ being assigned 0 and 1. These additions belong to horizontal separation since we added a new part to our system (the module linked to the light). One of these possible traces is depicted in Figure \ref{timelinewithx}. Some events are occurring simultaneously, for instance $t_{\text{on}}$ always occurs on an instant coincident to an occurrence of $t_{x_0}$. Such relations between events can be defined in \ccsl (a simple case of sub-clocking).

\begin{figure}[h]
  \centering
\begin{tikzpicture}[->,>=stealth',shorten >=1pt,auto,node distance=2.5cm,
                    semithick]
  \node[state]         (C)              {1};
  \node[state]         (D) [right of=C] {2};
  \node[state]         (E) [right of=D] {3};

  \node[initial,state] (A) [below of=C] {Off};
  \node[state]         (B) [below of=E] {On};

  \path (A) edge node {$t_\text{on}$}  (C)
        (B) edge node {$t_\text{off}$ $\{ x \leftarrow 0 \}$} (A)
        
        (C) edge node {$t_\text{stack}$}  (D)
        (D) edge node {$t_\text{compute}$}  (E)
        (E) edge node {$t_\text{store}$}  (B)
        
        (B) edge [loop right] node {$t_\text{ex}$ $\{ x \leftarrow 1 - x \}$} (B);
\end{tikzpicture}
\caption{The refined system}
\label{automaton3}
\end{figure}
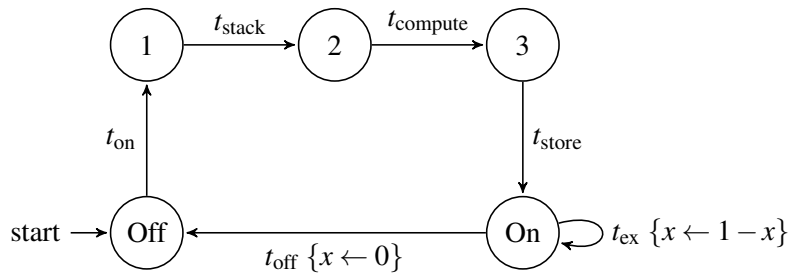

It is important to notice that when specifying the action executed by this system, we implicitly took a certain point of view. We deliberately ignored some lower level concerns such as the way a computer system handles a memory. This is where vertical separation takes place. Seeing closer to the machine will lead to other events which can refine the access to the variable $x$. For instance, the "switch on" event can be viewed as a succession of actions, such as powering up the system, retrieving the address of $x$, computing (here there is no actual computation since 1 is an atomic value, but there could be in the case of a more complicated expression) the value of $1$ and storing this value at the right address. These events, except for the first one, are used to handle the computation and the storing of a value in a memory. Taking into account these events require to view the system at a lower level than before, in which case its representation as a transition system is depicted in Figure \ref{automaton3}. 

\begin{figure}[h]
 
\begin{subfigure}{0.45\textwidth}

  \begin{tikzpicture}[axis/.style={thick, ->, >=stealth'},scale=1.1]
    \foreach \i in {1,...,5} {
      \draw[axis] (2,\getdata{\i}{y}) -- (6.5,\getdata{\i}{y}) node(xline)[right]{};
      \foreach \j in {1,...,3} {
        \draw[very thick] (\getdata{\j}{xTwo},\getdata{\i}{y}-0.1) --  (\getdata{\j}{xTwo},\getdata{\i}{y}+0.1);
      }
      \draw (1.2, \getdata{\i}{y}) node {\getdata{\i}{titles}};
    }
    \foreach \i in {1,...,3} {
      \draw [thick, dashed, Blue] (\getdata{\i}{xTwo},\getdata{1}{y}+0.3) -- (\getdata{\i}{xTwo},\getdata{5}{y}-0.3);
    }
    
     \foreach \i in {1,2} {
     	\draw [Red,thick,->, >=stealth'] (\getdata{\i}{xTwo},\getdata{1}{y}) -- 
     		(\getdata{\i+1}{xTwo},\getdata{5}{y});
     }    
    
  \end{tikzpicture}
  \caption{The higher level of observation}
  \label{Higherlevel}

\end{subfigure}
\begin{subfigure}{0.55\textwidth}

  \begin{tikzpicture}[axis/.style={thick, ->, >=stealth'},scale=1.1]
     \foreach \i in {1,...,5} {
       \draw[axis] (2,\getdata{\i}{y}) -- (8.2,\getdata{\i}{y}) node(xline)[right]{};
       \draw (1.2, \getdata{\i}{y}) node {\getdata{\i}{titles}};
     }
     \foreach \j in {1,...,3} {
       \draw[very thick] (\getdata{\j}{xOne},\getdata{1}{y}-0.1) --  (\getdata{\j}{xOne},\getdata{1}{y}+0.1);
     }
     \foreach \j in {1,...,12} {
       \pgfmathtruncatemacro\result{int(Mod(\j-1,4)+2)}
       \draw[very thick] (\getdata{\j}{xAllDots},\getdata{\result}{y} -0.1) --  (\getdata{\j}{xAllDots},\getdata{\result}{y}+0.1);
       \ifnum\j<12
       \pgfmathtruncatemacro\more{int(Mod(\j,4)+2)}
       \draw[Red,thick,->, >=stealth'] (\getdata{\j}{xAllDots},\getdata{\result}{y}) -- (\getdata{\j+1}{xAllDots},\getdata{\more}{y});
       \fi
     }
      
     \foreach \i in {1,...,12} {
       \draw [thick, dashed, Blue] (\getdata{\i}{xAllDots},\getdata{1}{y}+0.3) -- (\getdata{\i}{xAllDots},\getdata{5}{y}-0.3);
     }

    
  \end{tikzpicture}
  \caption{The lower level of observation}
  \label{Lowerlevel}
\end{subfigure}
\caption{Both levels of observation}
\label{bothlevels}
\end{figure}
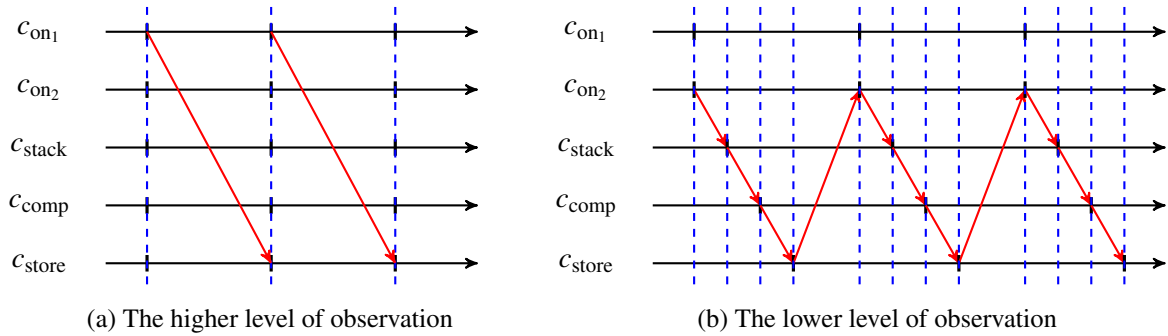

The "switch on" transition has been refined in several transitions.  $t_\text{on}$ represents the powering of the system, $t_\text{stack}$ the stacking of the address of $x$, $t_\text{compute}$ the computing of the value of the expression $1$ and $t_\text{store}$ the storing of the computed value at the stacked address. Note that we only refined one transition here for the sake of clarity and simplicity. Refining the other transitions would rely on exactly the same reasoning which is of no use for the relevance of this example. \\
\indent This analyse induces two different points of view on our system. The higher level of observation is represented on Figure \ref{Higherlevel}. The events that are not refined are omitted from now on, for the sake of clarity. They don't influence the reasoning we are conducting, thus their omission is acceptable.\\
\indent From the higher point of view, all the instants on which the sub-events occur are equivalent both to each other and to the containing event. Their underlying order is hidden and has no impact on the trace of the system at this level. The lower point of view, however, is different, as depicted on Figure \ref{Lowerlevel}.

\storedata{titles}{{$c_{\text{on}_1}$}{$c_{\text{on}_2}$}{$c_{\text{stack}}$}{$c_{\text{comp}}$}{$c_{\text{store}}$}}
\storedata{xOne}{{2.5}{4.5}{6.5}}
\storedata{xTwo}{{2.5}{4}{5.5}}

\storedata{y}{{1.4}{.7}{0}{-.7}{-1.4}}
\storedata{xAllDots}{{2.5}{2.9}{3.3}{3.7}{4.5}{4.9}{5.3}{5.7}{6.5}{6.9}{7.3}{7.7}}

For the lower level of observation, the different instants are ordered in a way such that they respect the specification in Figure \ref{automaton3}. The blue dashed lines represents the equivalence classes induced by the respective partial orders while the red arrows represent the precedent relations of these orders (we did not represent the links that can be deduced by transitivity or other properties of partial orders).

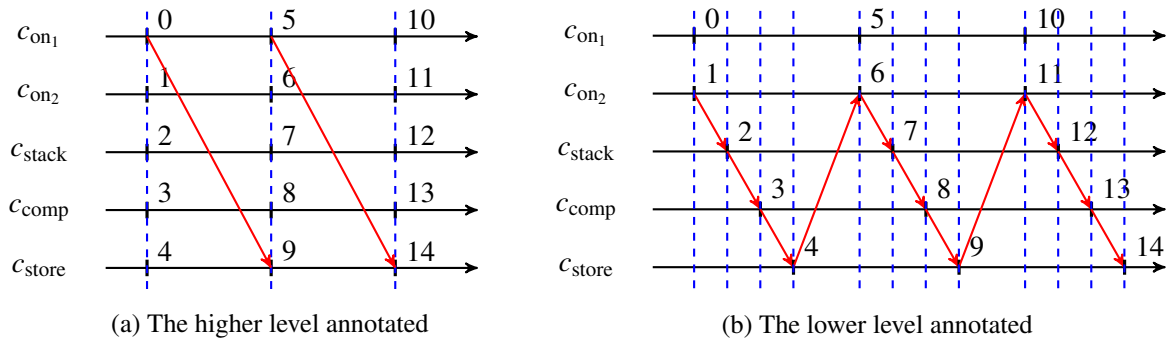
\begin{figure}[h]
 
\begin{subfigure}{0.45\textwidth}

\begin{tikzpicture}[axis/.style={thick, ->, >=stealth'},scale=1.1]
  \foreach \i in {1,...,5} {
    \draw[axis] (2,\getdata{\i}{y}) -- (6.5,\getdata{\i}{y}) node(xline)[right]{};
    \foreach \j in {1,...,3} {
      \pgfmathtruncatemacro\nodeNumber{int((\j-1)*5 + \i -1)}
      \draw[very thick] (\getdata{\j}{xTwo},\getdata{\i}{y}-0.1) coordinate (A)
      -- (\getdata{\j}{xTwo},\getdata{\i}{y}+0.1) coordinate (B);
      \node[above right=0.7mm and 0.1mm of A] {$\nodeNumber$};
    }
    \draw (1.2, \getdata{\i}{y}) node {\getdata{\i}{titles}};
  }  
  \foreach \i in {1,...,3} {
    \draw [thick,dashed, Blue] (\getdata{\i}{xTwo},\getdata{1}{y}+0.3) -- (\getdata{\i}{xTwo},\getdata{5}{y}-0.3);
  }
  
       \foreach \i in {1,2} {
     	\draw [Red,thick,->, >=stealth'] (\getdata{\i}{xTwo},\getdata{1}{y}) -- 
     		(\getdata{\i+1}{xTwo},\getdata{5}{y});
     }    
\end{tikzpicture}
\caption{The higher level annotated}
\label{Higherlevelannotated}

\end{subfigure}
\begin{subfigure}{0.55\textwidth}

  \begin{tikzpicture}[axis/.style={thick, ->, >=stealth'},scale=1.1]
     \foreach \i in {1,...,5} {
       \draw[axis] (2,\getdata{\i}{y}) -- (8.2,\getdata{\i}{y}) node(xline)[right]{};
       \draw (1.2, \getdata{\i}{y}) node {\getdata{\i}{titles}};
     }
     \foreach \j in {1,...,3} {
       \pgfmathtruncatemacro\nodeNumber{int((\j-1)*5)}
       \draw[very thick] (\getdata{\j}{xOne},\getdata{1}{y}-0.1) coordinate (A)
       -- (\getdata{\j}{xOne},\getdata{1}{y}+0.1) coordinate (B);
       \node[above right=0.7mm and 0.1mm of A]{$\nodeNumber$};
     }
     \foreach \j in {1,...,12} {
       \pgfmathtruncatemacro\modulo{int(Mod(\j,4))}
       \pgfmathtruncatemacro\result{int(Mod(\j-1,4)+2)}
       \coordinate (M) at (\getdata{\j}{xAllDots},\getdata{\result}{y});
       \node (C) [above=1mm of M] {};
       \node (D) [below=1mm of M] {};
       \draw[very thick] (C) -- (D);

       \node[above right=0.1mm and 0.1mm of M] {\theothers};
      
       \ifnum\j<12
       \pgfmathtruncatemacro\more{int(Mod(\j,4)+2)}
       \draw[Red,thick,->, >=stealth'] (\getdata{\j}{xAllDots},\getdata{\result}{y}) -- (\getdata{\j+1}{xAllDots},\getdata{\more}{y});
       \fi

       \stepcounter{others}
       \ifnum \modulo=0
       \stepcounter{others}
       \fi
     }
      
     \foreach \i in {1,...,12} {
       \draw [thick,dashed, Blue] (\getdata{\i}{xAllDots},\getdata{1}{y}+0.3) -- (\getdata{\i}{xAllDots},\getdata{5}{y}-0.3);
     }

    
  \end{tikzpicture}
  \caption{The lower level annotated}
  \label{Lowerlevelannotated}

\end{subfigure}

\caption{Both annotated levels of observation}
\label{bothannotated}
\end{figure}

Until now, the instants on which the events occur formed an unspecified set. Since our goal is to mechanize this example, we need to instantiate it to an actual set. We chose the natural numbers because they  allow to annotate the traces while expressing quite easily the relations at both levels of refinement. The annotated higher level of observation is given in Figure \ref{Higherlevelannotated}.

\newcounter{others}
\setcounter{others}{1}

This representation allows us to define the coincidence and the precedence relations that bind its different instants, as subsets of $\mathbb{N} \times \mathbb{N}$. Since both these relations must be transitive, the coincidence must be symmetrical and they must form a strict partial order. We omit the related elements which can be deduced from these properties.
\[\begin{array}{|C{1.9cm}|C{1.9cm}|C{1.9cm}|C{5.7cm}|}
\hline
\multicolumn{3}{|c|}{\textbf{Coincidence Relation}} & 
\textbf{Precedence Relation} \\
\hline
(0 , 1) & (0 , 2) & (0 , 3) & \multirow{2}*{(0 , 5)} \\ 
\cline{1-3}
(0 , 4) & (5 , 6) & (5 , 7) & \\
\hline
(5 , 8) & (5 , 9) & (10 , 11) &\multirow{2}*{(5 , 10)} \\
\cline{1-3}
(10 , 12) & (10 , 13) & (10 , 14) & \\
\hline
\end{array}\]

Since the traces are infinite, there are an infinite number of couples in each relations. We only expressed them for the visible subset. We now define these relations for any natural number, by relying on euclidean decomposition of their operands by 5: 
\vspace{.2cm}

\hspace{-.5cm}
\fbox{\parbox{.95\textwidth}{\begin{center}
\vspace{-.2cm}
$\forall (a,a') \in \mathbb{N}^2, \exists ! \text{ } (q,r,q',r') \in \mathbb{N}^4 \text{ } : a = 5q + r \wedge r < 5 \wedge a' = 5q' + r' \wedge r' < 5$
\vspace{-.2cm}
\end{center}
}}

\vspace{.2cm}

These relations, using the same notation, are defined as follow:

\vspace{.2cm}

\hspace{-.5cm}
\fbox{\parbox{.95\textwidth}{\begin{center}
\vspace{-.2cm}
$ \forall (a,a') \in \mathbb{N}^2, a \approx_2 a' \Leftrightarrow q = q'$ \\
$\forall (a,a') \in \mathbb{N}^2, a <_2 a' \Leftrightarrow q < q' $
\vspace{-.2cm}
\end{center}
}}

\vspace{.2cm}

The same work can be achieved for the lower level of observation, which is displayed on Figure \ref{Lowerlevelannotated}. The relations extracted from Figure \ref{Lowerlevelannotated} are depicted in the table below. As previously explained, only the relevant couples are mentioned. 
\[\begin{array}{|C{5.7cm}|C{1.9cm}|C{1.9cm}|C{1.9cm}|}
\hline
\textbf{Coincidence Relation} & \multicolumn{3}{|c|}{\textbf{Precedence Relation}} \\
\hline
(0 , 1) & (1 , 2) & (2 , 3) & (3 , 4) \\ 
\hline
(5 , 6) & (4 , 5) & (6 , 7) & (7 , 8) \\
\hline
(10 , 11) & (8 , 9) & (9 , 10) & (11 , 12)\\
\hline
\ldots & (12 , 13) & (13 , 14) & \ldots \\ 
\hline
\end{array}\]

By taking the same decomposition as before, we can mathematically define the relations at the lower level of observation.

\vspace{.2cm}

\hspace{-.5cm}
\fbox{\parbox{.95\textwidth}{\begin{center}
\vspace{-.2cm}
$ \forall (a,a') \in \mathbb{N}^2, a \approx_1 a' \Leftrightarrow (q_1 = q_2) \wedge ((r_1,r_2) \in [0,1]^2 \vee (r_1 = r_2 \wedge r_1 \notin [0,1]))$\\
$ \forall (a,a') \in \mathbb{N}^2, a <_1 a' \Leftrightarrow (q_1 < q_2) \vee ((q_1 = q_2) \wedge (r_1 < r_2) \wedge (r_2 \neq 1))$
\vspace{-.2cm}
\end{center}
}}

\vspace{.2cm}

Since both couples of relations have been defined mathematically, we can prove that they correspond to a situation of refinement. The proof has been done both on paper and in \agda, and is not presented here. It is however available on the first author's web page \footnote{\url{http://montin.perso.enseeiht.fr}}. The steps in this proof are the following:
\begin{itemize}
\item Prove that the two couples of relation form strict partial orders (12 predicates).
\item Prove that these orders satisfy the refinement relation (4 predicates).
\end{itemize}

\section{Mechanization of the refinement relation}

This work is supported by a significant effort of mechanization. We advocate that any proof and formalization should be done through formal methods in order both to ease and verify the mathematical content of the work. In our case, this effort has been done using a proof assistant called \agda. We briefly present it in this section before getting to the benefits of this mechanization. 

\subsection{\agda}

\agda is a dependently typed programming language developed by Ulf Norell at Chalmers University. As any other language, the types of which can depend on values, it is expressive enough to build mathematical theories, thanks to the Curry-Howard isomorphism, which ensures the correctness of any property whose equivalent type is inhabited. The core of the language is an intuitionist type theory, on which the well-known tool \coq is based as well. Although these two languages share the same heart, they are quite different when it comes to developing and proving properties. \coq uses named tactics, the action of which is hidden from the reader of the \coq file -- as well as the underlying lambda-terms -- while \agda provides a framework to help the programmer write them by hand, thus making them visible in the \agda file. This framework is what makes programming in \agda possible since typed lambda terms are arguably impossible to write without software assistance, assuming their type reaches a certain level of complexity.

\agda also differs from \coq by its native unification mechanism, which is usually summarized by "\agda allows to pattern-match on equality proofs". Although unification can hardly be reduced to this simple sentence, \agda indeed allows to case-split on the equality proofs, thus unifying the operands of the equality. More generally, \agda is able to infer, by unification, the value of variables present in the context of a proof. \coq does not provide such a straight-forward mechanism and handles cases usually solved by unification in \agda with other ways that we find less convenient. 

The rest of this paper contains small pieces of Agda code, depicting either data structures, predicates or proofs established during our development. Although these blocks help assessing the technical aspects of our work, their understanding is not mandatory to grasp the notions we describe and the reasoning behind them. Their goal is to briefly picture what \agda proofs look like and to help the reader assess the underlying effort of this work. Here is our relation written in \agda:
\begin{Verbatim}[fontsize=\scriptsize, frame=single, framesep=2mm]
_≺≈_ : ∀ {ℓ} → Rel (Rel A ℓ × Rel A ℓ) _
(_≈₁_ , _≺₁_) ≺≈ (_≈₂_ , _≺₂_) =
  (∀ {a b} → a ≺₁ b → a ≺₂ b ⊎ a ≈₂ b) × 
  (∀ {a b} → a ≺₂ b → a ≺₁ b) ×
  (∀ {a b} → a ≈₂ b → a ≈₁ b ⊎ a ≺₁ b ⊎ b ≺₁ a) × 
  (∀ {a b} → a ≈₁ b → a ≈₂ b)
\end{Verbatim}

\subsection{Properties of the refinement relation}

It looks reasonable to assume that our refinement relation should be a strict partial order between strict partial orders. Being able to prove such property would enforce the correctness of our definition towards the refinement requirement. However, as we mentioned earlier, a strict partial order is based on an equivalence relation. This relation could be the propositional equality, or another relation that we defined. We tried both possibilities, and we present the results of these attempts in this section.

\subsubsection{It is a pre-order towards propositional equality}

As a reminder, a pre-order is an algebraic structure composed of an equivalence relation and a precedence relation which is transitive and reflexive according to the equivalence relation. We showed that our refinement relation formed a pre-order towards the propositional equality. The propositional equality, in dependent types, is a family of types generated by the reflexivity rule. This means that two quantities are propositionally equal if they were built with the same constructors. We start by proving that our relation is transitive:
\begin{Verbatim}[fontsize=\scriptsize, frame=single, framesep=2mm]
trans≺≈ : ∀ {ℓ} → Transitive (_≺≈_ {ℓ})
trans≺≈ p q {a} {b} with p {a} {b} | q {a} {b} | p {b} {a}
trans≺≈ p q {a} {b} | pr₁ , pr₂ , pr₃ , pr₄ | pr₅ , pr₆ , pr₇ , pr₈ | _ , pr₁₀ , _ , _
  = (λ x → case pr₁ x of (λ {(inj₁ x₁) → pr₅ x₁ ; (inj₂ y) → inj₂ (pr₇ y)})) ,
    (λ x → pr₂ (pr₆ x)) , (λ x → pr₇ (pr₃ x)) , (λ x → case pr₈ x of (λ {(inj₁ x₁) → pr₄ x₁ 
    ; (inj₂ (inj₁ x₁)) → inj₂ (inj₁ (pr₂ x₁)) ; (inj₂(inj₂ y)) → inj₂ (inj₂ (pr₁₀ y))}))
\end{Verbatim}

We also prove it is reflexive:
\begin{Verbatim}[fontsize=\scriptsize, frame=single, framesep=2mm]
refl≺≈ : ∀ {ℓ} → Reflexive (_≺≈_ {ℓ})
refl≺≈ = (λ x → inj₁ x) , (λ x → x) , (λ z → z) , (λ x → inj₁ x)
\end{Verbatim}

This allows us to exhibit the pre-order we aimed for.
\begin{Verbatim}[fontsize=\scriptsize, frame=single, framesep=2mm]
preorder≺≈≡ : ∀ {ℓ} → IsPreorder _≡_ (_≺≈_ {ℓ})
preorder≺≈≡ = record { isEquivalence = isEquivalence
  ; reflexive = λ {i} {j} x → case x of (λ {refl → refl≺≈ {x = i}}) ; trans = trans≺≈ }
\end{Verbatim}

\subsubsection{It is a partial order towards the equivalence between relations}

Two relations are equivalent when the subset they form are equal. We implemented this definition for our couples of relations:
\begin{Verbatim}[fontsize=\scriptsize, frame=single, framesep=2mm]
_≈≈_ : ∀ {ℓ} → Rel (Rel A ℓ × Rel A ℓ) _
(_≈₁_ , _≺₁_) ≈≈ (_≈₂_ , _≺₂_) = ∀ {a b} →
  (a ≈₁ b → a ≈₂ b) × 
  (a ≈₂ b → a ≈₁ b) × 
  (a ≺₁ b → a ≺₂ b) × 
  (a ≺₂ b → a ≺₁ b)
\end{Verbatim}

A partial order is a pre-order with an anti-symmetrical property between its two underlying relations. We already proved that our refinement relation was transitive, but we still need to prove that our equivalence relation is indeed an equivalence and that the properties of reflexivity and antisymmetry hold between them. The equivalence is obvious and is not presented here. The reflexivity is proved as follow:
\begin{Verbatim}[fontsize=\scriptsize, frame=single, framesep=2mm]
refl≺≈≈ : ∀ {ℓ} → (_≈≈_ {ℓ}) ⇒ (_≺≈_ {ℓ})
refl≺≈≈ x {a} {b} with x {a} {b}
refl≺≈≈ x {a} {b} | proj₃ , proj₄ , proj₅ , proj₆ =
  (λ x₁ → inj₁ (proj₅ x₁)) , (λ x₁ → proj₆ x₁) , (λ x₁ → proj₃ x₁) , (λ x₁ → inj₁ (proj₄ x₁))
\end{Verbatim}

As for the antisymmetry:
\begin{Verbatim}[fontsize=\scriptsize, frame=single, framesep=2mm]
antisym≺≈ : ∀ {ℓ} → Antisymmetric _≈≈_ (_≺≈_ {ℓ})
antisym≺≈ x x₁ {a} {b} with x {a} {b} | x₁ {a} {b}
antisym≺≈ x₁ x₂ {a} {b} | proj₃ , proj₄ , proj₅ , proj₆ | proj₇ , proj₈ , proj₉ , proj₁₀
  = (λ x → proj₅ x) , (λ x → proj₉ x) , (λ x → proj₈ x) , (λ x → proj₄ x)
\end{Verbatim}

This allows us to exhibit the partial order we aimed for:
\begin{Verbatim}[fontsize=\scriptsize, frame=single, framesep=2mm]
partialOrder≺≈≈ : ∀ {ℓ} → IsPartialOrder _≈≈_ (_≺≈_ {ℓ})
partialOrder≺≈≈ = record 
  { isPreorder = record { isEquivalence = equiv≈≈ ; reflexive = refl≺≈≈ ; trans = trans≺≈ }
  ; antisym = antisym≺≈ }
\end{Verbatim}

\subsection{It preserves \ccsl operators}

{\setlength{\parindent}{0cm}
\textbf{\ccsl denotational semantics:} In a previous work, we mechanized the denotational semantics of \ccsl in \agda. This section gives the required notions about this mechanization in order to connect it to our refinement relation.}

\ccsl is based on clocks, which represents the different occurrences of a specific event. Typically, a clock represents one of the different timelines we depicted in the different figures in this paper. In our work, we represent clocks by a record containing a predicate to emulate the subset of instants on which this clock ticks, and a predicate which makes sure the ticks of the clocks are totally ordered regarding the given strict partial order:
\begin{Verbatim}[fontsize=\scriptsize, frame=single, framesep=2mm]
record Clock : Set₁ where
  constructor
    clock
  field
    Ticks  : Pred Support lzero
    TicTot : _≺_ isTotalFor Ticks
\end{Verbatim}

\ccsl provides several constructs to constrain the different clocks of a system amongst each other. They are grouped into two different categories: the relations and the expression. A relation is a relation between two clocks, while an expression, in our denotational semantics, is a relation over three clocks:

\vspace{-.5cm}

\begin{multicols}{2}
\begin{Verbatim}[fontsize=\scriptsize, frame=single, framesep=2mm]
Relation : Set₁
Relation = Clock → Clock → Set
\end{Verbatim}
\begin{Verbatim}[fontsize=\scriptsize, frame=single, framesep=2mm]
Expression : Set₁
Expression = Clock → Clock → Clock → Set
\end{Verbatim}
\end{multicols}

\vspace{-.3cm}

Since this paper is not meant to detail the whole semantics, we only give one example for each of these categories. The relation we present is the sub-clocking. A clock $c_1$ is a sub-clock of a clock $c_2$ when $T(c_1) \subset T(c_2)$:
\begin{Verbatim}[fontsize=\scriptsize, frame=single, framesep=2mm]
_⊑_ : Relation
(clock Tc₁ _) ⊑ (clock Tc₂ _) = ∀ {x₁} → x₁ ∈ Tc₁ → ∃ \x₂ → x₁ ≈ x₂ × x₂ ∈ Tc₂
\end{Verbatim}

The expression we will present is the union. A clock $c$ is considered the union of a clock $c_1$ and a clock $c_2$ when $T(c) = T(c_1) \cup T(c_2)$:
\begin{Verbatim}[fontsize=\scriptsize, frame=single, framesep=2mm]
_≡_∪_ : Expression
clock Tc _ ≡ clock Tc₁ _ ∪ clock Tc₂ _ =
    (∀ {i} → (Tc₁ i ⊎ Tc₂ i) → ∃ \j → i ≈ j × Tc j)
    × (∀ {i} → Tc i → ∃ \j → i ≈ j × (Tc₁ j ⊎ Tc₂ j))
\end{Verbatim}

{\setlength{\parindent}{0cm}
\textbf{A relation between clocks:} This clock definition allows extending our refinement relation to clocks. Informally, a clock refines another one when it represents a thinner event which was hidden by the first clock. For instance, if we get back to our example, the "switch on" clock is refined by several clocks, including the "compute" one. Let us consider the following definition:}
\begin{Verbatim}[fontsize=\scriptsize, frame=single, framesep=2mm]
_refc_ : REL (Clock _) (Clock _) _
(clock Ticks₁ _) refc (clock Ticks₂ _) =
  × (∀ {x} → Ticks₂ x → ∃ λ y → Ticks₁ y × (y ≈₂ x))
  × (∀ {x} → Ticks₁ x → ∃ λ y → Ticks₂ y × (y ≈₂ x))
\end{Verbatim}
A clocks refines another if they are defined on refined partial orders, while also obeying the following predicates: each tick of the more abstract clock is refined by at least one tick of the concrete clock and each tick of the concrete clock is the refinement of a tick of the abstract clock. 

\vspace{.3cm}

{\setlength{\parindent}{0cm}
\textbf{Proofs of semantic preservation:} We prove the preservation of the semantics of the \ccsl constructs towards the refinement relation. This preservation is described and discussed about the two semantic elements we presented. The proofs are not presented here because they are not relevant but they are available on-line. The preservation property about sub-clocking is as follows: given four clocks $c_a, c_b, c_1, c_2$, if $c_1$ is a sub-clock of $c_2$, if $c_1$ refines $c_a$ and $c_2$ refines $c_b$ then $c_a$ is a sub-clock of $c_b$.}
\begin{Verbatim}[fontsize=\scriptsize, frame=single, framesep=2mm]
subclockingRefinement : c₁ ⊑₁ c₂ → c₁ ≺refc c₁₁ → c₂ refc c₂₂ → c₁₁ ⊑₂ c₂₂
\end{Verbatim}

The preservation property about union is a follows: given four clocks $c_0$ $c_1$, $c_2$ and $c$, if $c_1$ refines $c$, if $c_2$ refines $c$ and if $c_0 = c_1 \cup c_2$ then $c_0$ refines $c$.
\begin{Verbatim}[fontsize=\scriptsize, frame=single, framesep=2mm]
unionRefinement : c₁ refc c → c₂ refc c → c₀ ≡ c₁ ∪ c₂ → c₀ refc c
\end{Verbatim}

\section{Conclusion}

\subsection{Assessment}

This contribution provided a refinement relation for time models in order to allow system developers to focus on their own view of the system rather than a common one shared among all of them. This enables seizing their constraints without taking into account considerations from other levels of observation. The constraints on the system can then be described and solved at all different levels with the assurance that none of them will be compromising the others. Furthermore, refinement can be used to specify simulations and bi-simulations between systems. In this case, the two specifications are either different ways of specifying its behaviour or different systems that must satisfy the same interface.

More precisely, this paper presented a relation over strict partial orders, the goal of which is to model instant refinement. Each level of observation is represented by a specific strict partial order. All of these orders must be  proven bond through our notion of refinement in order to ensure the correctness of the whole system. This work is mechanized in \agda, and is connected to a mechanization of \ccsl developed using the same tool. This allowed us to formally prove different algebraic properties about refinement itself as well as the preservation of several \ccsl operators through our relation of refinement.

\subsection{Future works}

Several future works are currently being conducted:
\begin{itemize}
\item[$\bullet$] We would like to allow the system developers to express their constraints at the most suitable level of abstraction. This could only be done if their constraints are propagated in the other levels where other constraints are specified. Thus, we plan to complete the link between \ccsl and our instant refinement through the proof of multiple preservation properties. This extension could also lead to automated reduction of constraints sets relying on additional properties about \ccsl operators.
\item[$\bullet$] The \ccsl team at INRIA\footnote{We thank the \ccsl team at INRIA for the fruitful discussions we had around \ccsl and the need for a refinement relation.} plans to integration our notion of refinement to their toolsets. While refinement cannot be considered as yet another \ccsl operator, it could still be used to provide more expressiveness through the use of several partial orders instead of a single one. 
\item[$\bullet$] We would like to apply our approach to a more complex example. In that purpose, we would like to refactor and complete our previous work regarding the proof of correctness of a translation between langages (process models to Petri nets) as a weak bisimulation that binds these languages with our refinement relation. Indeed, weak bisimulation can be seen as a special case of refinement.
\end{itemize}

\vspace{-.5cm}

\bibliographystyle{eptcs}
\bibliography{bibrefine}

\end{document}